\documentclass[prl, showpacs, twocolumn, floatfix]{revtex4}

\usepackage{graphicx}
\usepackage{amsmath, amsfonts, amssymb, bm}
\usepackage[]{psfrag}

\psfragscanoff
\begin{document}

\title{Microscopic laser-driven high-energy colliders}
\author{Karen Z. \surname{Hatsagortsyan}}
\email{k.hatsagortsyan@mpi-hd.mpg.de}
\author{Carsten \surname{M\"{u}ller}}
\email{c.mueller@mpi-hd.mpg.de}
\author{Christoph H. \surname{Keitel}}
\email{keitel@mpi-hd.mpg.de}
\affiliation{Max-Planck-Institut f\"ur Kernphysik, Saupfercheckweg 1, D-69117 Heidelberg, Germany}

\date{\today}

\begin{abstract}
The concept of a laser-guided $e^+e^-$ collider in the high-energy regime is presented and its
feasibility discussed. Ultra-intense laser pulses and strong static magnetic fields are employed to 
unite in one stage the electron and positron acceleration and their head-on-head collision. We show 
that the resulting coherent collisions in the GeV regime yield an enormous enhancement of the 
luminosity with regard to conventional incoherent colliders.
\end{abstract}
 

\maketitle

Since the invention of laser light amplification with chirped pulses, extremely short and strong 
laser fields have been generated with ever increasing intensities \cite{CPA}. 
With current laser devices reaching $10^{22}$ W/cm$^2$, electron acceleration to hundreds
of MeV \cite{accel_el} has been realized along with clear evidence for laser-induced nuclear 
reactions \cite{Ledingham}.
Following \cite{Mourou}, the feasible increase of laser beam sizes from today's $1$ cm to $1$ m and of
 the laser pulse energy from $1$ J to $10$ kJ should enhance the focused intensity by further four
 orders of magnitude to $10^{26}$ W/cm$^2$. 
Proposals for reaching even higher laser fields have been put forward lately \cite{Pukhov} aiming at 
the intensity threshold $I\approx 10^{29}$ W/cm$^2$ for $e^+e^-$ pair production in plain vacuum. 
Thus there are clear prospects for high energy and particle physics with laser pulses rather than 
conventional accelerators.

In this letter feasible collider schemes to realize particle physics with high-power lasers are 
investigated. Rather than improving current laser accelerator schemes, we show how to unite more 
naturally acceleration and focusing with laser-guided {\it coherent} head-on-head collisions.
For the example of an $e^+e^-$ collider, emphasis is placed on optimizing the collision
geometry, i.e. maximizing the center-of-mass (c.m.) energy, and minimizing the dispersion of the 
rapidly propagating quantum wave packets.

Particle physics applies to distances of order $r \sim 1\ {\rm fm} = 10^{-13}$ cm and below, where the 
strong interaction comes into play. This corresponds to energies $\varepsilon \sim c\hbar/r \sim 1$ 
GeV, with speed of light $c$ and Planck constant $\hbar$.
Hence, a total $e^+e^-$ c.m. energy of more than $\varepsilon \sim 1$ GeV is required 
to probe strong interaction, e.g. via quark-antiquark creation, where the cross-section resonantly 
enhances at $\varepsilon \sim 4$ GeV. Alternatively the weak interaction can be probed via $Z^0$
($\varepsilon >91$ GeV) or $W^{+}W^{-}$ production ($\varepsilon >162 $ GeV), while there are also 
mainly electromagnetic processes such as $\mu^+ \mu^-$ ($\varepsilon >0.212$ GeV) and $\tau^+ \tau^-$ 
($\varepsilon >3.56$ GeV) creation in this energy regime (see Fig.1).  Those processes are addressed 
in what follows with lasers, refraining thus from the TeV regime (and luminosities of $10^{34}$cm$^{-2}$s$^{-1}$) aspired to go beyond the standard model \cite{collider}.

\begin{figure}[ht]
\begin{center}
\includegraphics{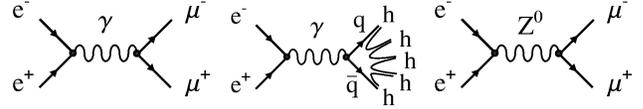}
\caption{\label{mmqqZ0}Diagrams representing examples of particle creation processes that can be 
realizable by the laser-driven colliders discussed here: $\mu^+\mu^-$, $q\bar{q}$ (with further 
evolution into hadrons $h$) or $Z^0$ production (with further decay into $\mu^+\mu^-$).} 
\end{center} 
\end{figure}

\begin{figure}[b]
\vspace{-0.5cm}
\includegraphics{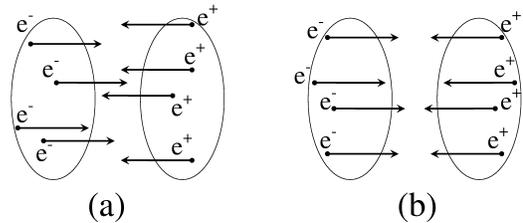}
\vspace{-0.4cm}
\caption{\label{microcollis} (a) In conventional $e^+e^-$ colliders bunches of accelerated $e^+$ and $e^-$ 
are focused to collide head-on-head {\it incoherently}, i.e., bunches collide head-on-head but electrons 
and positrons in the bunch do not. (b) With laser fields, $e^+$ and $e^-$ originating from Ps may 
collide head-on-head {\it coherently}.} 
\end{figure}

Conventional $e^+e^-$ colliders \cite{accel} consist of two distinct acceleration and focusing devices.
After their acceleration, bunches of $e^\pm$ are focused by magnetic lenses and brought into head-on-head
collision. However, the particles in the bunch are distributed randomly such that each microscopic $e^+e^-$ 
collision is not head-on-head but has a mean impact parameter $\bar{\rho}_{incoh}$ determined 
by the beam radius $r_b$: $\bar{\rho}_{incoh}\sim r_b$, characterizing the collision as incoherent 
(see Fig.\ref{microcollis}a). In principle, 
the acceleration stage could also be realized by some laser-based technique, like the 
wakefield acceleration \cite{wakefield}, as applied lately also to positrons \cite{wakefield+}. 
Here care has to be taken for $e^\pm$ phase synchronism, focusing would remain a separate stage 
and the collisions incoherent.

Our system of interest is laser-driven positronium (Ps) where, due to the equal constituent masses and
the oscillating nature of laser fields, acceleration and focusing can be realized in one single stage. 
As main advantage and contrary to conventional colliders, this laser-driven collider (LDC) provides coherent 
head-on-head $e^+e^-$ collisions (see Fig. \ref{microcollis}b) in contrast to 
the incoherent collisions described above. The reason for this is that in the LDC the colliding $e^+$ and 
$e^-$ stem from the same Ps atom, i.e., their initial coordinates are confined within an atomic size of the
order of the Bohr radius $a_B$. Moreover, both the $e^+$ and $e^-$ are coherently driven by the same laser 
field such that the mean impact parameter at the coherent recollision is microscopic, i.e. 
$\bar{\rho}_{coh}\sim a_B$. A further advantage is, that the energy-momentum conservation is fulfilled 
for photon absorptions without requiring any additional interaction. In fact, when the $e^\pm$ 
annihilate at the collision {\em inside} the laser field they absorb energy, temporarily acquired from the 
laser field before. 

The coherent head-on-head collisions can improve the luminosity significantly. The luminosity $\mathcal{L}$ 
is the accelerator physics parameter that determines the number of reaction events $\mathcal{N}$ per unit 
time \cite{accel}, i.e. $\frac{d\mathcal{N}}{dt}= \sigma \cdot \mathcal{L}$ with $\sigma $ being the reaction 
cross-section. In the case of colliding $e^\pm$ bunches with a random distribution of the particles in the 
bunch we have as luminosity $\mathcal{L}_b=(N_{e}^2/S_b)f$. Here $N_{e}$ is the number of particles in 
the bunch, $S_b$ the bunch cross-sectional area and $f$ the bunch repetition frequency. The above 
formula means that each of the $N_e$ particles in the bunch yields $N_{e}\sigma /S_b$ reaction events.
In the case of coherent head-on-head collisions the luminosity $\mathcal{L}_m$ has a coherent component as 
each coherent collision yields $\sigma/a_{\bot}^2$ reaction events:
\begin{equation}
\mathcal{L}_m=\left[\frac{N_{e}(N_e-1)}{S_b}+\frac{N_{e}}{a_{\bot}^2}\right]f \label{L}
\end{equation}
where $a_{\bot}$ is the particle's wave-packet transversal size with respect to the collision momentum.
If the spreading of the particle's wave packet is rather small 
\begin{equation}
N_ea_{\bot}^2<S_b ,\label{micro}
\end{equation}
then we can speak of a coherent head-on-head collision. Coherent collisions are especially helpful when the 
initial densities of the $e^+$ and $e^-$ (Ps atoms) are low. Let us illustrate the typical scales: If the 
size of the particle bunch, which should not be larger than the waist size of the laser beam, is  
$l_b\sim 3 \; \mu $m, then $S_b\sim 10^{-7}$ cm$^2$ (and the bunch volume $V_b\sim 10^{-10}$ cm$^3$). 
If now the electron wave packet at the collision has spread to a size not larger than $a_{\bot} \sim 10~\AA $, 
then for $N_e\sim 10$ (corresponding to an electron density of $n_e\sim 10^{11}$ cm$^{-3}$) the luminosity 
is $10^{6}$ times enhanced due to its coherent component.

Single laser pulses are little suitable to realize an LDC because i) the main part of the laser energy is 
misused to accelerate the created particles along the laser beam direction and ii) the spreading of the 
$e^+e^-$ wave packet is large, deluting the advantage of coherent collisions. Consider an $e^-$ (or $e^+$) 
moving in a strong laser field with $\xi:=eA_0/mc^2\gg 1$, where $A_0$ is the amplitude of the laser's 
vector-potential and $e$ and $m$ are the $e^-$ charge and mass, respectively. Then the $e^-$ longitudinal 
momentum along the laser propagation direction $p_z\sim mc\xi^2/2$ considerably exceeds the transversal 
one $p_x\sim mc\xi$ \cite{Landau}. That is, the $e^-$ acquires a large total energy from the laser field, 
while only a small part of it ($\sim p_x/p_z\sim 1/\xi$) is connected with the transverse momentum and is 
thus transformed into the rest mass of the created particles. For $\xi=2000$, e.g., which is realized at 
$I=8\cdot 10^{24}$W/cm$^2$ and a laser wavelength of $\lambda =0.8 \mu$m, the total energy lies in the 
TeV range, whereas only 2 GeV can be used for mass production. The rest of the energy just serves to 
accelerate the created particles. Further, due to the large drift velocity of the $e^-$, its oscillation 
period and, thus, the recollision time in the $e^+e^-$ c.m. frame are much larger than the laser period. 
This dramatically increases the electron wave-packet spreading, which can be estimated within the 
following intuitive picture. In the $e^+e^-$ c.m. frame, the initial momentum spread $\delta p^{\prime}$ of 
the electron wave-packet is determined by its initial spatial size of order of the Bohr radius $a_B$, 
that is $\delta p^{\prime}\sim \hbar/a_B$. Then the spreading in this (primed) frame during the recollision 
time $t_{r}^{\prime}$ can be estimated as 
\begin{equation}
\delta x^{\prime}\sim \delta z^{\prime} \sim \delta p^{\prime}t_{r}^{\prime}/m\sim c \alpha t_{r}^{\prime},
\label{spreading}
\end{equation}
with the laser polarization and propagation directions $x$ and $z$, respectively, and the fine-structure 
constant $\alpha$. Since the recollision time at $\xi \gg 1$ equals $t_{r}^{\prime}\sim 4\pi \gamma_z/\omega$, 
with the $\gamma$-factor of the c.m. frame $\gamma_z\sim \xi$ and the laser frequency $\omega$, 
the spreading in this frame equals $\delta x^{\prime}\sim \delta z^{\prime}\sim \xi\alpha\lambda$ with
the laser wavelength $\lambda$. 
Hence, at the collision point the wave-packet size is of the order of a conventional particle beam 
size for $\xi\sim 1000$ and the advantage of coherent collisions is removed. 

\begin{figure}[h]
\includegraphics{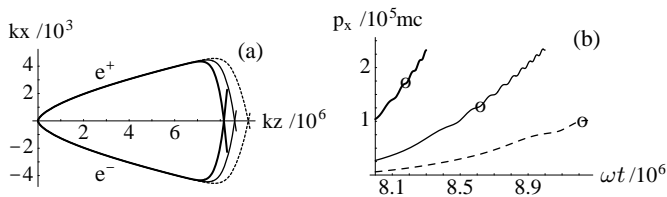}
\caption{\label{kick} Electron dynamics in the fields of a laser pulse and a spatially confined 
magnetic field. The laser field is linearly polarized with an intensity parameter of $\xi =3000$ 
($I=1.9\cdot 10^{25}$W/cm$^2$ at $\lambda =0.8 \, \mu$m).
The static magnetic field is directed along the laser magnetic field and spatially confined within 
a Gaussian centered at $kz_0=9\cdot 10^6$ ($k=2\pi/\lambda$) and with a width $d$ of $kd=10^6$. The 
thick, thin, and dashed lines correspond to the magnetic field magnitudes $B_0=10^4$ T, $3\cdot 10^3$ T, 
and $10^3$ T, respectively. (a) shows the $e^+e^-$ trajectories.
(b) shows the time $t$ dependence of the electron transversal momentum $p_x$. The recollision points are 
indicated by a circle (o): $\omega t_r/10^6=8.17, 8.65$ and $9.23$, for the above mentioned 
magnetic fields.}
\end{figure}

One way of improving the relative orientation of $e^+$ and $e^-$  at the scattering moment is the 
application of an additional static magnetic field to the laser pulse. In the magnetically assisted 
collision setup the $e^\pm$ first acquire a large longitudinal momentum from the laser field, which 
then is turned into the transversal direction by a magnetic kick, boosting this way the $e^+e^-$ 
scattering energy in their c.m. frame.
Rather than the less favorable uniform field, a spatially confined magnetic field was employed.
For the parameters in Fig.\ref{kick} a total scattering energy of 100 GeV ($\varepsilon \sim  35mc^2\xi$) 
is achievable, being almost 35 times larger than without magnetic field and sufficient to produce
 a $Z^0$ boson. The efficiency defined as the ratio of the energy used for particle production 
to the total energy is substantially enhanced as compared to the case of the single laser pulse: the 
longitudinal momentum exceeds the transversal one only by a factor of $100\ll\xi$. Increasing the magnetic 
field does not lead to a further improvement of the efficiency.  

\begin{figure}[b]
\includegraphics{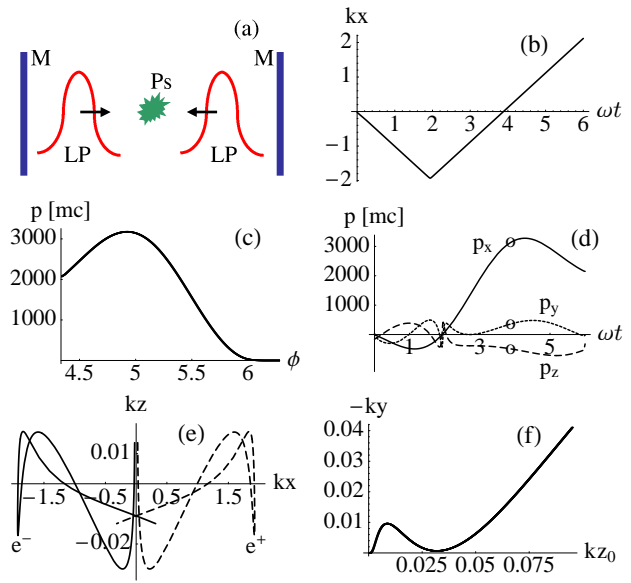}
\caption{\label{crossed} Gas of Ps atoms in counterpropagating, circularly polarized laser pulses (LP) 
with $\lambda =0.8 \mu$m and $\xi =1000$ ($I=2\cdot 10^{24}$W/cm$^2$). The pulses are cos$^2$-shaped, 
with sin-oscillations and have a steeply rising edge.
The initial position of Ps is $x_0={\rm y}_0=0,kz_0=0.05$, the initial velocity zero. 
(a) setup of the process with mirrors $M$.
(b) time dependence of the electron transversal coordinate with rapid recollision time $\omega t_r=3.9$.
(c) electron momentum at the recollision as a function of the carrier-envelope phase of the laser pulse.
(d) time dependence of the electron momentum components for $\phi=5$ rad, where 
$p_{\rm y}$ and $p_z$ are enhanced by a factor of 20 and the recollision point is marked by circles. 
(e) $e^\pm$ trajectories in the $xz$-plane.
(f) electron y-offset at the recollision as a function of the initial $z_0$-offset.}
\end{figure}

\begin{figure}[h]
\includegraphics{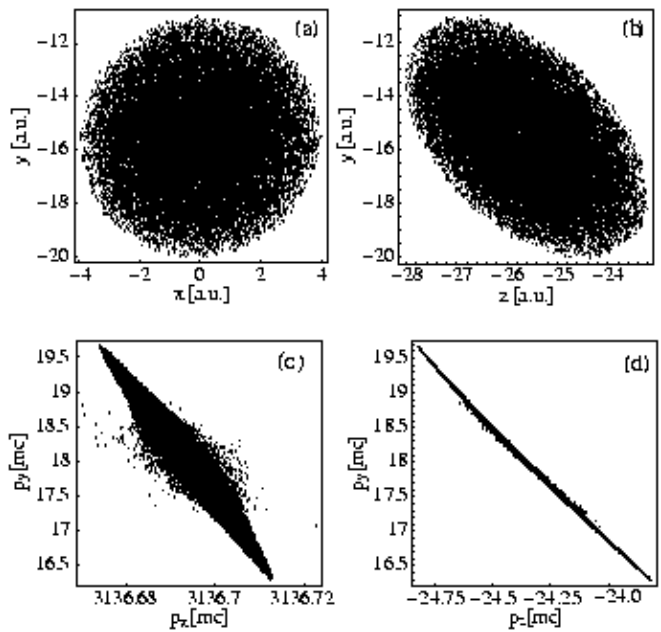}
\caption{\label{MCcrossed} Monte-Carlo simulation \cite{Monte-Carlo} of the $e^+e^-$ wave-packet dynamics 
in counterpropagating, circularly polarized laser pulses for the same parameters as in Fig. 4. 
(a)-(d): Coordinate- and momentum-space distributions of one of the particles at the recollision.  
The initial $z$-offset ($kz_0=0.05$) leads to a y-offset at the recollision, such that the average 
impact parameter is $\bar{\rho}\approx 15\AA$ (rather than $\bar{\rho}\approx 1\AA$ for $kz_0=0$).}
\end{figure}

Although the magnetic kick enhances scattering energy and efficiency significantly, the latter 
is still not high. The relativistic red shift has lead to a long acceleration phase and, along with
the resulting high energy, unfortunately also to a large spreading and thus reduced collision efficiency. 
Replacing the magnetic field by a second counterpropagating laser pulse of the same intensity (see Fig. \ref{crossed}a) will cancel 
the drift and with this the enlarged acceleration time. Moreover, if the laser beams are circularly polarized
and equal handed, the resulting magnetic field, being directed along the electric field, induces a focusing 
force and the wave-packet spreading is substantially reduced \cite{corkum}. The $e^\pm$ dynamics in the 
counterpropagating laser pulses, calculated on base of the classical equations of motion, is shown in Fig. \ref{crossed}. In the $z=0$ symmetry plane the resulting field has practically only an $x$-component. The 
recollision time $t_r$ is defined by the condition $x=0$ (see Fig. \ref{crossed}b). The suppression of the 
electron drift induces rapid head-on-head recollisions during one laser period in the lab frame. 
We note, however, that the $e^+$ and $e^-$ positions at $t_r$ might differ in their y-coordinates. By a 
suitable choice of the absolute phase of the two laser waves one can simultaneously achieve small 
y-offset and large momentum at the collision (see Fig. \ref{crossed}c). Then, the $e^-$ momentum in 
the laser field has mainly an $x$-component ($p_x\approx 3100 mc$) and a total c.m. energy of $3.2$ GeV is
reached (see Fig. \ref{crossed}d). The y- and $z$-components of the scattering momentum depend 
on the initial location of the Ps atom along the beam axis, but are always orders of magnitude smaller: 
$p_{\rm y}, p_z \lesssim 10\,mc$ for $kz_0\lesssim 0.05$ (see also Fig. \ref{MCcrossed}d). Thus, there is little 
laser energy misdirected ($\lesssim 1 \%$) though the collision energy is still relatively small because of
the short acceleration periods. The $e^\pm$ trajectories in the $xz$-plane are shown in Fig. \ref{crossed}e.
The distance along the y-axis between the recolliding particles also depends on the initial $z$-coordinate 
of the Ps atom (see Fig. \ref{crossed}f). To keep the impact parameter below ca. $10\AA$, 
the initial offset from the symmetry plane should not exceed $kz_0\approx 0.06$. 

Due to the magnetic focusing effect, the spreading of the electron wave-packet in the counterpropagating laser 
pulses is under control, as can be infered from Fig. \ref{MCcrossed} where the recolliding wave-packets are shown. 
In the $x$y-plane there is practically no spreading as the wave-packet radius of about 4 a.u. coincides with its initial value (see Fig. \ref{MCcrossed}a). In $z$-direction the magnetic field even leads to a compression 
of the wave-packet (see Fig. \ref{MCcrossed}b), which is the stronger the closer the Ps initially was to the symmetry plane. The small wave-packet size allows to fulfill the coherent collision condition in Eq.\,(\ref{micro}). The crucial parameter determining the collisional impact parameter is thus the $e^+e^-$ distance along the y-axis 
as shown in Fig. \ref{crossed}f. The momentum distributions show that the particles have a rather well-defined 
collision momentum that is almost parallel to the $x$-axis (see Figs. \ref{MCcrossed}c, d).

Regarding achievable luminosities, the typical cross-section of particle creation processes via 
$e^+e^-$ collisions (see Fig.1) is of the order 
$\sigma_{typical}\sim \pi r_{e}^{2}\gamma^{-2}\sim 10^{-30}$ cm$^{-2}$, where $r_e$ is 
the classical $e^-$ radius and $\gamma$ the collisional $\gamma$-factor. If we assume that $10^6$ 
scattering events per running year ($\sim10^7$ s) are sufficient for statistics: 
$\mathcal{N}=\sigma \int \mathcal{L}dt=10^6\; {\rm events/year}$, then the required
luminosity is ${\cal L}=10^{29}$ cm$^{-2}$s$^{-1}$. Thus, taking advantage 
of coherent collisions [see Eq.(\ref{L})] with $N_e=10^4$, the repetition rate shall be 
$f=10^{11}$ s$^{-1}$. This will require the supply of Ps atoms with a rate of 
$\frac{dN_{Ps}}{dt}=N_ef\sim 10^{15}$ s$^{-1}$. 
As an alternative, for a $e^+e^-$ plasma the luminosity is 
$\mathcal{L}_b=n_{e}^2fV_bl_b$ with plasma density $n_e$. 
For $V_b=10^{-9}$cm$^3$, $l_b=10^{-3}$ cm, and $f\sim t_{r}^{-1}\sim 10^{15}$s$^{-1}$, a luminosity 
of $\mathcal{L}_b=10^{29}$cm$^{-2}$s$^{-1}$ is achieved at  $n_e=10^{13}$ cm$^{-3}$. For comparison, 
present experimental data for $e^+$ plasmas give $n_e\sim 4\times 10^{9}$ cm$^{-3}$ \cite{surko} 
but one has to take into account that a new alternative laser-based technique is emerging with a 
recent proposal aiming at $n_e\sim 10^{16}$ cm$^{-3}$ \cite{wilks}.

Concluding, two laser-driven $e^+e^-$ colliders are put forward and compared in Table I with a conventional 
scheme. An LDC based on a gas of Ps atoms in the field of two crossed laser beams will enable 
a high scattering luminosity by using coherent $e^+e^-$ head-on-head collisions. Because of the
rapid acceleration phase the attainable energy of this scheme is constricted by the energy of order 
$mc^2\xi$ acquired by the $e^\pm$ along a laser wavelength. Considerably higher energies can be 
attained in the setup that combines a single laser field with a static magnetic kick but at the 
price of a substantialy larger wave-packet spreading.
To circumvent this problem, an $e^+e^-$ plasma rather than a Ps gas was considered in Table I. 
Dependent on the particular need, high energy or high efficiency, LDC1 or LDC2 is preferable, respectively.

Funding by Deutsche Forschungsgesellschaft via KE-721-1 and Alexander-von-Humboldt foundation
for KZH is acknowledged. We thank B. Henrich for help at the onset of this project.

\begin{table}[h]
\begin{tabular}{|l||c|c|c|}\hline
 Parameter  
 &\ \ LDC1\ \ 
 &\ \ LDC2\ \
 &\ \ LEP \ \   \cr\hline\hline
 Collision energy (GeV)               & 10$^2$    & 10$^1$     & 10$^2$     \cr
 Spatial extension (cm)               & 10$^0$    & 10$^{-4}$  & 10$^6$     \cr 
 Particles per bunch                  &           & 10$^4$     & 10$^{11}$  \cr 
 $e^+e^-$ plasma density (cm$^{-3}$)  & 10$^{13}$ &            &            \cr
 Repetition rate (s$^{-1}$)           & 10$^{15}$ & 10$^{11}$  & 10$^5$     \cr
 Luminosity (cm$^{-2}$s$^{-1}$)\ \  & 10$^{29}$ & 10$^{29}$  & 10$^{32}$  \cr\hline
\end{tabular}\\
\caption{Comparison of the proposed LDC (at $\lambda =0.8\; \mu$m, $\xi=3000$) with the conventional 
high-energy $e^+e^-$ collider ring LEP at CERN. LDC1 denotes the magnetically assisted setup, 
LDC2 the crossed-beams configuration.}
\end{table}

\end{document}